\documentclass[conference]{IEEEtran}
\usepackage[numbers]{natbib}
\usepackage{amsmath,amssymb,amsfonts}
\usepackage{algorithmic}
\usepackage{graphicx}
\usepackage{textcomp}
\usepackage{xcolor}
\usepackage{hyperref}
\usepackage{csquotes}
\begin{document}

\title{Managing Risk in DeFi Portfolios}

\author{\IEEEauthorblockN{Hugo Inzirillo}
\IEEEauthorblockA{\textit{CREST - Institut Polytechnique de Paris}\\
\textit{Simons}\\
Email: hugo@simons.finance}
\and
\IEEEauthorblockN{Stanislas De Quénetain}
\textit{BEX - Blockchains Expert}\\
\IEEEauthorblockA{\textit{Stan}\\
Email: stan@stan.finance}
}

\maketitle

\begin{abstract}
Decentralized Finance (DeFi) is a new financial industry built on blockchain technologies. Decentralized financial services have consequently increased the ability to lend, borrow, and invest in decentralized investment vehicles, allowing investors to bypass third party intermediaries. DeFi's promise is to reduce the cost of transaction and management fees whilst increasing trust between agents of the Financial Industry 3.0. This paper provides an overview of the different components of DeFi, as well as the risks involved in investing through these new vehicles. We will also propose an allocation methodology which will integrate and quantify these risks.  
\end{abstract}
\IEEEpeerreviewmaketitle
\section{Introduction}

In the context of the expansion of Decentralized Finance Companies \cite{CHEN2020e00151}, to better understand the concept behind Decentralized Finance (DeFi), it is first necessary to have an understanding of its structure and its constitutive elements. In DeFi, yields are generated using 4) tokens deposited into 3) liquidity pools generated by 2) protocols hosted on 1) blockchains. We will start by clarifying these terms. 

A blockchain is the database of a network. Its main purpose is to record the transactions that have been realized among all members of the network. For instance, the Bitcoin network is composed of all of the people who have downloaded the Bitcoin source code (generally through a wallet) and the blockchain will then record every Bitcoin that is sent from one person to another. It is important to note that the Bitcoin source code is very basic in the sense that it can only be used to receive and send Bitcoins. Other networks, such as Ethereum, have a more complex source code. Which, for example, can be used to send and receive Ether (the Ethereum native token), but can also read programs called smart contracts. So within the Ethereum blockchain, you can save transactions used to send Ether, but also save smart contracts and the transactions that are actioning these smart contracts. In other words, Ethereum (like other smart contract blockchains such as Solana, Terra, BSC...) will rent its blockchain (that stores smart contracts and transactions) and its virtual machine (required to read and execute smart contracts) to projects that want to use a decentralized environnement without dealing with the burden of maintaining a blockchain infrastructure. 

\begin{figure}
    \centering
    \includegraphics[width=1\columnwidth]{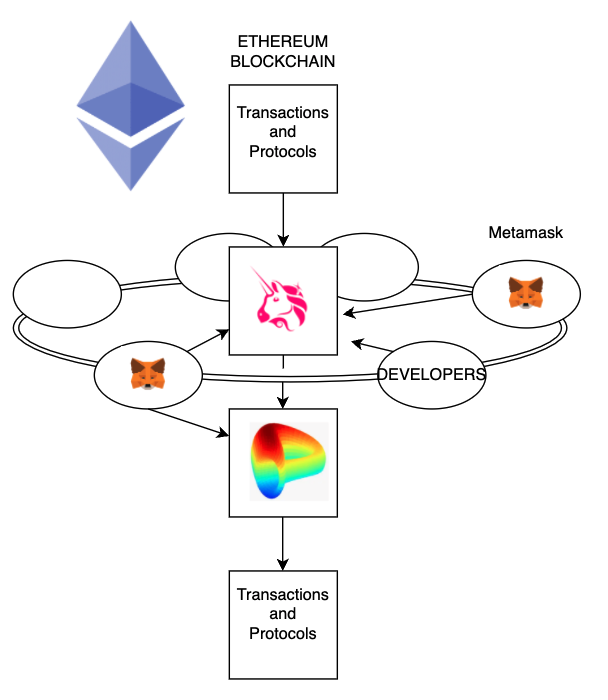}
    \caption{DeFi Structure on Ethereum}
    \label{fig:defi_structure}
\end{figure}

Let us take the example of Uniswap, the most famous decentralized exchange (\enquote{DEX}) of Ethereum. The founders of the project have written a source code (a smart contract) and have decided to host Ethereum in order to benefit from Ethereum's decentralized architecture, security, and its important network of miners and users. The purpose of the smart contract is to create liquidity pools that will allow users to swap tokens, we will explain this in more detail later. Concretly, the smart contract is recorded in a block of the Ethereum blockchain and is actioned through transactions. For instance, every time a person wants to swap two tokens in a given pool, he will send a transaction from his wallet (e.g., Metamask). This transaction will action a function of the smart contract and the new state of the smart contract (e.g., the new amount of tokens in the pool) will be recorded on the blockchain. As such, more operations are possible on the Ethereum blockchain than are possible on the Bitcoin blockchain. 

In our example, Uniswap is a protocol, also called a DApp or Decentralized Application. This application is hosted on Ethereum and will also be able to generate its own token (i.e., the \enquote{UNI}). It has its own business logic and tokenomics which are not related to Ethereum's own tokenomics. 

\begin{figure}
    \centering
    \includegraphics[width=1\columnwidth]{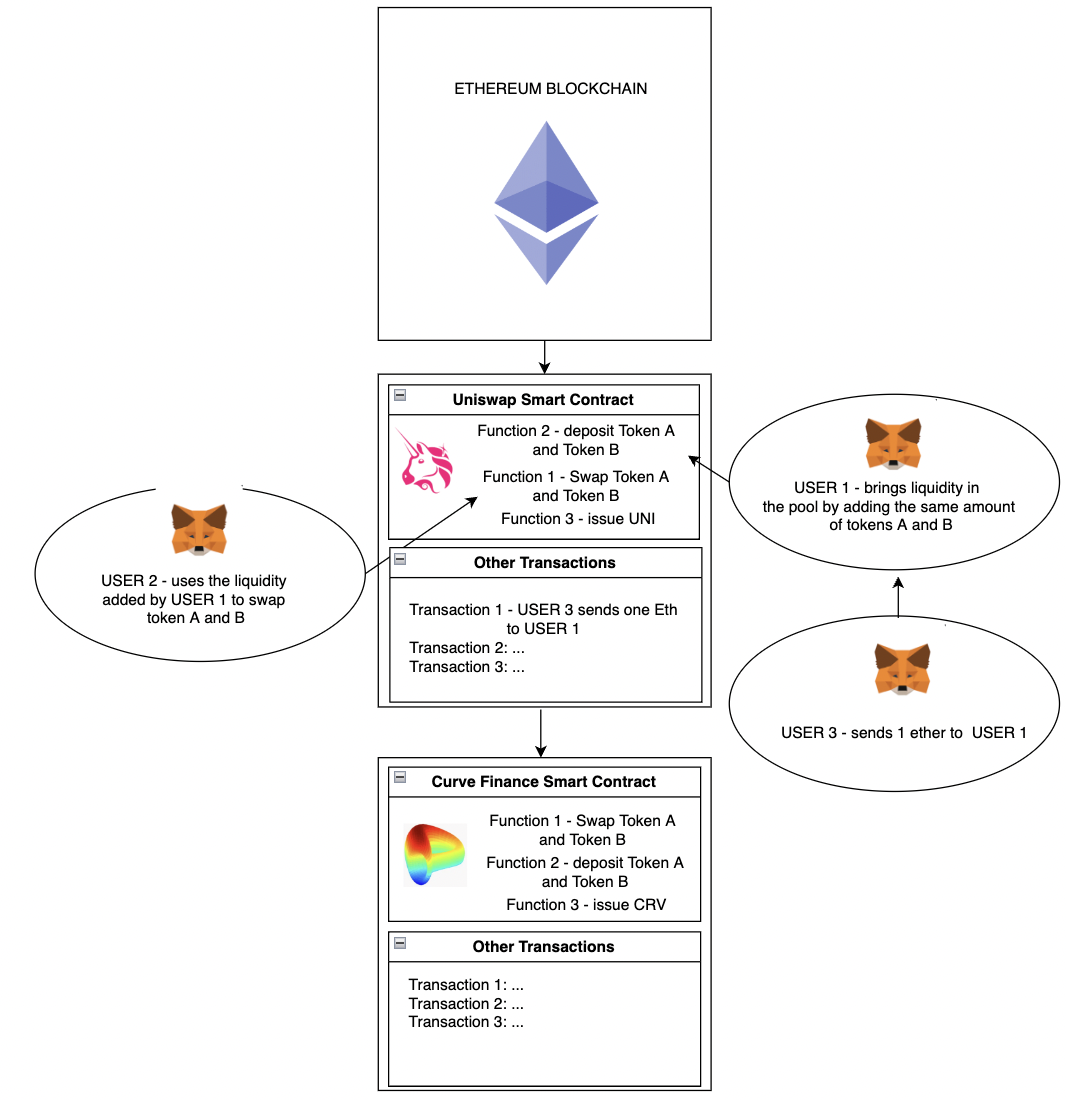}
    \caption{DeFi DApps on Ethereum}
    \label{fig:DApp}
\end{figure}

Uniswap founders will only pay a fee (i.e., a \enquote{Gas Fee}) to Ethereum when the source code is uploaded (i.e., \enquote{compiled}) on the blockchain for the first time and every time it is subsequently updated thereafter. Users of Uniswap will also pay Gas Fees every time they will action a function of this smart contract. Every time you realise a swap or deposit liquidity on Uniswap, you action a function of the smart contract, hence this is why you pay Gas Fees. Given the popularity of Ethereum, storage and execution of smart contracts is currently extremely expensive on Ethereum. This is why this part is increasingly delegated to Layer 2 such as Polygon, Arbitrum, or Optimism. 

\section{Decentralized Investment Vehicles}
\subsection{Staking vs Liquidity Mining}
Before we investigate how liquidity pools and liquidity mining work, it is important to clarify how different they are from the concept of \enquote{Staking}. These terms are regularly confused with eachother, but they are each referring to entirely different areas. The term \enquote{Staking} appears with decentralized protocols developing \enquote{Proof of Stake} consensus mechanisms for their mining processes. The purpose of consensus mechanisms with decentralized applications relying on a blockchain, is to secure the network by efficiently selecting the miners who will add the transactions to the blocks, subsequently add them to the blockchain, and then obtain the corresponding rewards.  The \enquote{Proof of Stake} like \enquote{Proof of Work}, or the more widely known \enquote{Delegated Proof of Stake}, are different types of consensus mechanisms which offer different methods of securely selecting miners. In terms of \enquote{Proof of Stake}, miners need to pledge a certain amount of the native tokens they initally bought in order to be eligible to add blocks to the blockchain. The idea behind this mechanism is that the more stake a miner has in a projet (i.e., the more tokens he has pledged in a protocol), the less interest he has to act viciously, as he will otherwise loose those tokens. So the term \enquote{Staking} refers to the pledging of tokens to a protocol by a miner in order to be eligible according to the consensus mechanism. 

The term \enquote{Liquidity Mining} is rather complex, as it refers to the mining process without having anything to do with it. The reference to \enquote{mining} has been added because the concept is similar. The \enquote{Liquidity Provider} that we will describe further below, will also pledge his token in the protocol but for a different purpose. The main differences are that 1) he is not necessarily a miner, 2) he will be able to pledge several tokens 3) that might not be the protocol native token  4) in a liquidity pool, 5) in order to obtain a trading fee. 

In order to explain how a liquidity pool works, we will now describe all these differences. 

The purpose of a liquidity pool is to act as a vault in which \enquote{Liquidity Providers} can pledge one or several tokens so that the \enquote{Traders} can use these tokens. Liquidity pools are used in DeFi for very different reasons, such as swapping, lending, borrowing, derivatives, and insurance. All the services of traditional finance are able to be proposed without intermediaries through these liquidity pools. 

In practice, a liquidity pool is indeed only a smart contract hosted on a decentralized blockchain such as Ethereum, Solana or Terra. Since these blockchains are open networks, anyone disposing of a Web3 wallet (Metamask for instance) is able to connect to the liquidity pool in order to deposit tokens or to use the service provided by the pool. Anyone can become a liquidity provider or a trader and interact directly with one another without an intermediary. 

We will now describe how the liquidity pool shown in Fig. \ref{fig:DApp} allows the traders to use all the tokens pledged in the pool by the liquidity providers in order to exchange them (or \enquote{swap} them). Let's imagine that a liquidity pool is dedicated to USDC and USDT. The liquidity provider will deposit the same amount of both tokens in the pool. In exchange, he will receive a \enquote{Liquidity Provider Token} or \enquote{LP Token} which corresponds to his share in the pool. If the pool contains 100 USDC/ 100 USDT and the liquidity provider deposits 10 USDC/10 USDT, he will detain 10 percent of the pool and receieve a corresponding LP token. In other words, the LP tokens are the proof of the deposit he made into the pool and he will need this proof in order to withdraw  his tokens at any point. 

\subsection{Liquidity Pools}

\begin{figure}
    \centering
    \includegraphics[width=1\columnwidth]{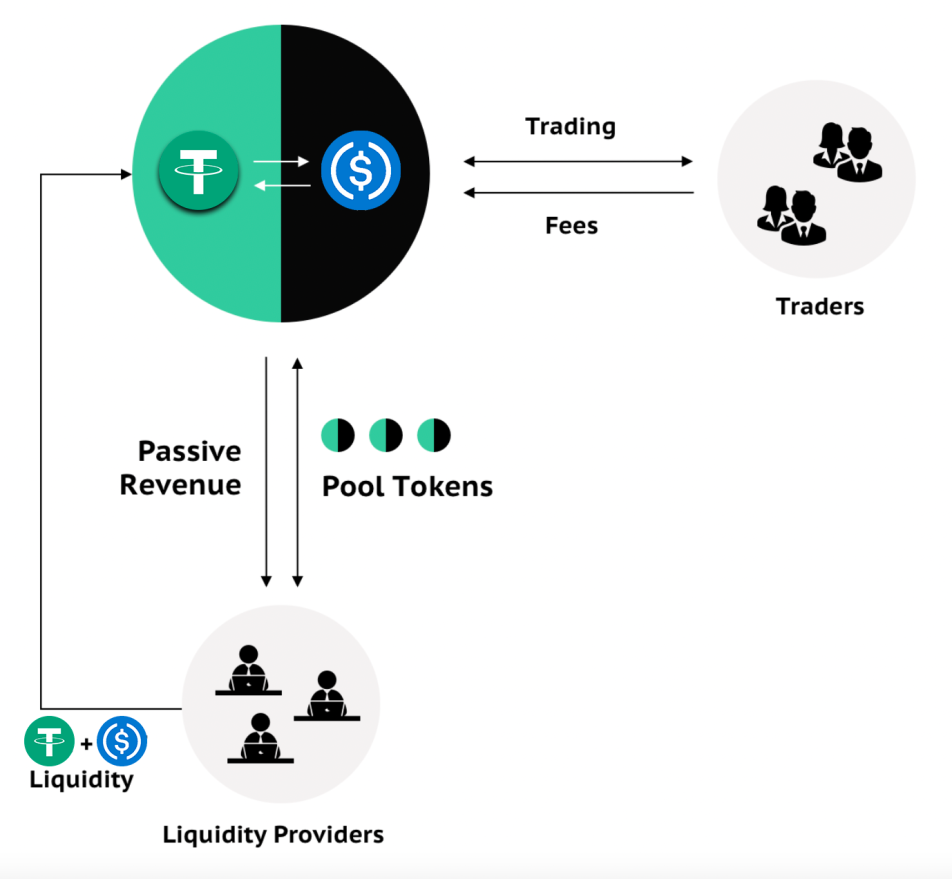}
    \caption{Liquidity Pool Flows}
    \label{fig:lp}
\end{figure}

Fig. \ref{fig:lp} describes how a liquidity pool actually works. 

The trader will use the pool in order to exchange USDT for USDC, or vice versa, and will pay a small fee for this service. The fee is almost entirely returned to the liquidity provider in proportion to his share in the pool. Hence the importance of LP tokens. When the liquidity provider wants to withdraw his USDT and USDC, he will send his LP token back to the protocol as a proof of detention. It is also possible to use this LP token on other protocols to obtain additional rewards. This is called \enquote{Yield Farming}.

\section{Risk Assessment in Decentralized Finance}
As just mentioned, blockchains are hosting protocols that are smart contracts. These smart contracts are creating liquidity pools in which tokens are deposited in order to generate liquidity for traders, and generate yields for the liquidity providers. 

So there are four different layers (i.e., blockchain, protocol, pool, token), each of them involving their own risks. \cite{jensen2020managing} identified risks associated with software integrity and security in DeFI.

\subsection{Blockchain Risks}

Describing in detail the risks that a decentralized network relying on a blockchain faces, would need more than a single paper. The general idea is that the blockchain as a database of the network needs to ensure that the transactions realized among the members of the network are added in a certain order, and that this order cannot be changed. 
Each of the blocks of a blockchain are used to timestamp all the transactions which prevents the occurence of a bitcoin being spent twice. The \enquote{double spending} attack is one of the most common attacks in a decentralized network, and one of the most difficult to solve. 

In the bitcoin network for instance, it is only possible to revert on a block which has been added to the blockchain (and thus double spend one of the transactions included in the block), by detaining more than 51\% of all the hashrate available on the network. The hashrate is the mining power available on the network at any given time. Even if one entity were to obtain 51\% of the hashrate, which has already occurred with one of the main mining pools in 2014, it would be far too expensive and an economic non-sense to recomplete the Proof of Work of the blocks that have already been mined in order to access a transaction. 

This discussion is not in the scope of this paper, but it is important to understand that the authenticity and the order of the data added to the blockchain is paramount for decentralized networks. For this reason, the consensus mechanisms in place need to be strong enough to ensure that miners with ill-intent do not compromize the entire network. 

With Proof of Work, the security and the selection of the miners is done through the cost of mining, as just explained. In other words, it is too expensive to revert on a block and double spend a transaction. 

With Proof of Stake networks, which will be used by Stan, miners are incentivized to behave honestly through a different method. As explained above, in order to be eligible to participate in the mining process, i.e., to add transactions in blocks and add blocks to the blockchain, they need to \enquote{stake} (i.e pledge) a certain amount of tokens native to the chain. If a miner behaves with ill-intent, he loses his stake. Generally speaking, the more miners that are dedicated to a network, the more competition you have among miners, and thus the more secure a network is. 

It does not mean that the source code of blockchains cannot be hacked, but this is an unlikely occurance for the main network, at least for the ones that have been selected by Stan.

\subsection{Protocol, Liquidity Pool, Token Risks}

Protocols are smart contracts that can be represented by state
machines \cite{protocols} (i.e programs) that are stored in the blocks of a blockchain and actioned through transactions.
In DeFi, most of these smart contracts create liquidity pools that could be considered a single point of failure and all tokens are indeed locked in these open source programs. Altough smart contracts are protected cryptographically, it is common to see hackers finding breaches in these smart contracts and exploiting them in order to steal the tokens of a pool. Since the appearance of DeFi, these hacks have been very common. This is why investing in DeFi requires a very careful selection of the protocols that are to be used. The allocation strategy must also be designed to spread the risk as much as possible in accordance with the risk of each protocol. 

It is important to note that tokens could also be a point of failure. For instance, certain types of stablecoins are considered more risky than others. This risk should also be taken into consideration in the allocation model. For simplicity reasons, the token's risks will not be inserted in the example we are proposing for this paper. 

It is also important to note that the contagion of failure works in different ways. A hack on a protocol will very rarely impact the blockchain on which it is hosted or impact the tokens that are sorted in a pool of liquidity. In other words, the tokens will be stolen i.e., moved to a public address that belongs to hackers. However, the code of the smart contract that generated these tokens will very rarely be impacted, but the hack of a protocol could impact other protocols especially due to \enquote{Yield Farming}.

\subsection{Yield Farming Risks}

\enquote{Yield Farming} refers to the use of different protocols in order to increase the returns on one given token. It is common to talk about \enquote{yield farming strategies}. 
For instance, a simple strategy could be to deposit bitcoins (wrapped as an ERC20) on Compound.finance as collateral and borrow stablecoins that will then be deposited on Uniswap for which you will receive an LP token which will possibly be optimized on Beefy.finance or Autofarm. In other words, you will remain exposed to Bitcoin while generating rewards on Uniswap and Beefy. This is a very simple example of a farming strategy and it can become much more complicated or leveraged, and is thus out of the scope of this paper.

We wanted to highlight that with yield farming, protocols are getting increasingly more connected over time. In our example, a hack on Beefy.finance would have an impact on Uniswap and Compound.finance. 
This risk needs to be taken into account when allocating tokens on different protocols. 
\section{Risk Parity Allocation Model}
Managing portfolios using blockchain technology is a new research area for the financial industry. The objective remains the same as that observed in the traditional asset management industry. However, risk diversification approach differs because there are new risks inherent to blockchain technology. These risks can be identified, but there is not yet a well established methodology to measure them. Stan.finance and Simons.finance have worked closely to develop a scoring methodology for blockchains and protocols. However, the literature on modeling is not very rich. In this paper we propose a risk parity approach based on the DeFi protocol scoring methodology. Portfolios can be managed by type of management (i.e., active or passive) or by risk profile. Portfolios are generally built over one metric: the risk/return ratio. Asset managers will look forward to generating higher returns for a defined risk established among the asset allocation.

\subsection{Risk Matrix}
Since the risk associated with DeFi is different from that of traditional financial assets, it is appropriate to construct a dedicated risk budget matrix. Here, the risk score retained is a score that aggregates the criteria of robustness, security, and reliability of a given protocol.The i-th protocol will be denoted by $P_i$, then $\sigma_{p}^{(i)}$ will be the risk score of the i-th protocol. 
Let us denote  $\sigma_p \triangleq \{ \sigma_{p}^{(1)},..., \sigma_{p}^{(n)}  \} $ as our vector of risk score for each protocol, hence our risk matrix will be denoted by $\Sigma$ and defined as:

\begin{equation}
    \Sigma = \text{Diag}(\sigma_p),
    \label{eq:raw_risk_score}
\end{equation}

where $\Sigma \in \mathbb{R}_{+}^{n \times n}$. The risk score used in equation \ref{eq:raw_risk_score} is obtained from a rating agency. We do not adjust the score. The objective is to propose a portfolio management style. From our risk score matrix we will normalize the scores as follows:

\begin{equation}
    \Tilde{\Sigma} = \frac{\Sigma}{||\Sigma||} 
\end{equation}

\subsection{Weighting Methodology}

Let us first define an Equal-risk Contribution (ERC) portfolio such a portfolio where the risk contribution $\sigma_i(w)$ is the same for all the assets
where $w$ is the weight matrix such $w \in W$. Note that $w^{*} = \{w \in {\big[0,1\big]}^n, \mathbf{1}w=1\}$. 
Let $\sigma$ be a continuously partially
differentiable risk measure. From the definition of marginal and total risk contributions \cite{Maillard60} we define this risk measure as:

\begin{equation}
    \sigma(w) = \frac{\partial \sigma(w) }{\partial w} = \sum_i^{n} w_i \frac{\partial \sigma(w) }{\partial w}  = \sum_i^{n}\sigma_i(w),
\end{equation}
and we aim to build a portfolio as such:

\begin{equation}
\label{eq:erc_equality}
     w_{i}^{*} \frac{\partial \sigma(w)}{\partial w_i} \Bigr|_{\substack{w=w^{*} }}= w_{j}^{*}\frac{\partial \sigma(w)}{\partial w_j} \Bigr|_{\substack{w=w^{*} }}
\end{equation}

As mentioned, we look for a combination  where the risk contribution of the protocols we invest in are equal. From equation \ref{eq:raw_risk_score} we obtain $\sigma_i(w^*)=\sigma_j(w^*)$. The objective is then to minimize the variance of the risk contribution, 

\begin{equation}
\begin{split}
    \underset{w\in W}{\mathrm{min}} & = f(w)\\
    u.c &
    \left\{
\begin{array}{rcr}
\mathbf{1}^T w = 1 \\
w \in [0;1]
\end{array}
\right.
\end{split}
\end{equation}

where  $f(w)$ is the variance of risk distribution defined as:

\begin{equation}
    f(w) = \sum_{i=1}^n \sum_{j=1}^n  ( w_i(\Tilde{\Sigma}w)_i - w_j(\Tilde{\Sigma}w)_j )^2
\end{equation}

under the fully invested and long only constraint.

\section{Application}
We introduced a methodology to allocate funds in protocols by considering the associated risk of each protocol. In this section, we will apply this methodology with market data and therefore simulate portfolios. First we obtained the data from the DeFi via API and then applied some transformation to build our DeFi portfolio.
For the experiment we built three portfolios: one that was weighted by the TVL; another Equally-Weighted (EW); and finally an ERC portfolio.
\subsection{Data}
 We  built DeFi portfolios using available yields in DeFi and selected the  project according to our internal scoring methodology. Within the protocol universe we used a score for each protocol that allowed us to measure the risk of our allocation, compute simulations for portfolios, and to apply the FX effect USDc/USD.

\subsection{Portfolios}
For the application we chose five protocols and computed a score using a discretionary methodology. We then selected the top APY available for stablecoins. We decided to include only stablecoins to avoid the price fluctuation of the token stacked in the protocol. 

If we used other tokens we would have applied the fx performance for each token against the base currency (e.g., USD). The idea is to draw the backtest of portfolios and compare the allocation models in terms of performance and risk.
Among the weighting methodology we built our risk ajusted vector. To do so, used our internal scoring methodology. \subsubsection{Equal-Weighted Porfolios}
We allocated the same percentage of stablecoin to each pool and computed the daily APY of our portfolio. The portfolio is rebalanced every day at midnight UTC for simplicity.

\begin{figure}[h!]
 \centering
    \includegraphics[width=1\columnwidth]{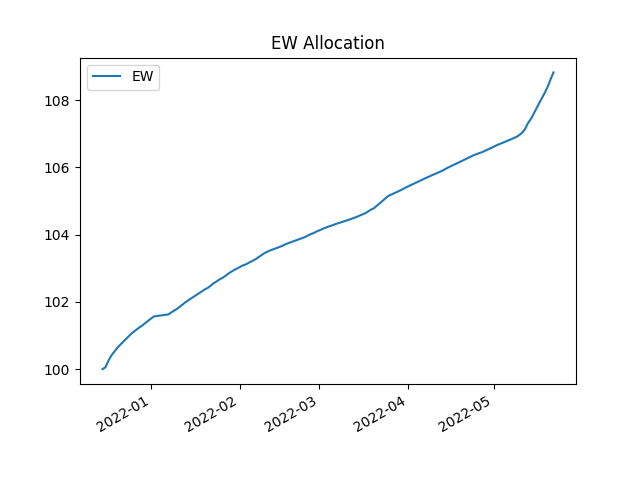}
    \caption{DeFi portfolio USDc based}
    \label{fig:port_ew_usdc}
\end{figure}

In order to consider an investor interested in depositing some USD, we applied a changing rate for stablecoin to obtain a USD based backtest. We observed some fluctuation.\\

\begin{figure}[h!]
    \centering
    \includegraphics[width=1\columnwidth]{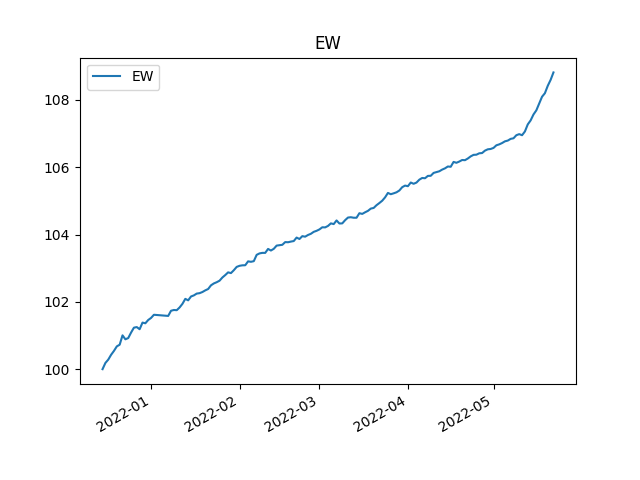}
    \caption{DeFi portfolio USD based}
    \label{fig:port_ew_usd}
\end{figure}

\begin{figure}[h!]
    \centering
    \includegraphics[width=1\columnwidth]{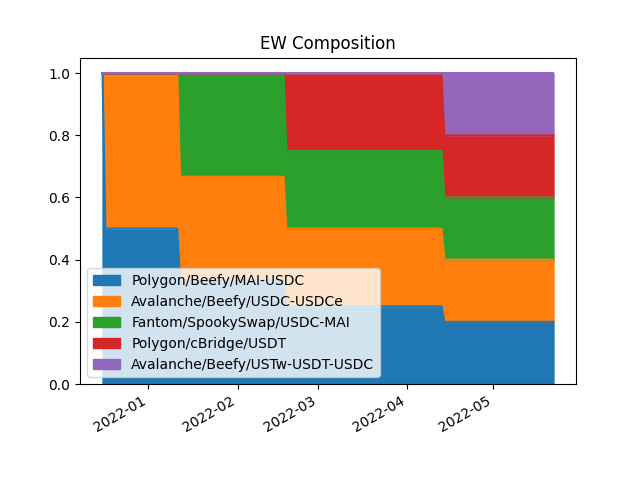}
    \caption{DeFi portfolio Composition}
    \label{fig:port_ew_compo}
\end{figure}

\subsubsection{Equal-Risk Contribution Portfolio}
We built an allocation using the risk scores to have the same proportion of risk contribution on each protocol and computed the daily APY. The ERC will rebalance the portfolio so that the risk contribution of each protocol is the same on a daily basis. 

\begin{figure}[h!]
 \centering
    \includegraphics[width=1\columnwidth]{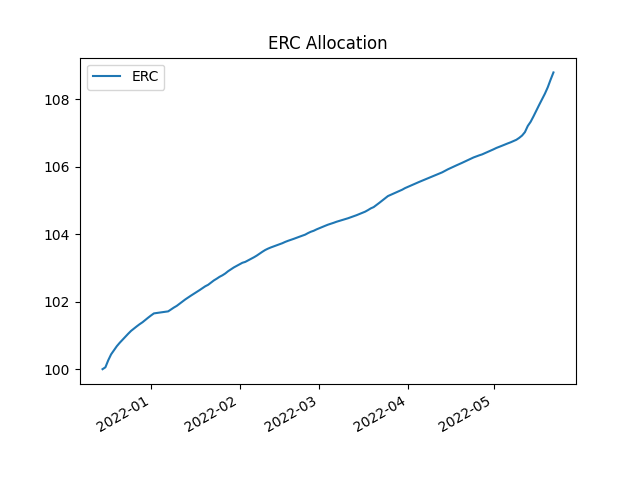}
    \caption{DeFi portfolio stablecoin based}
    \label{fig:port_erc_usdc}
\end{figure}

\begin{figure}[h!]
    \centering
    \includegraphics[width=1\columnwidth]{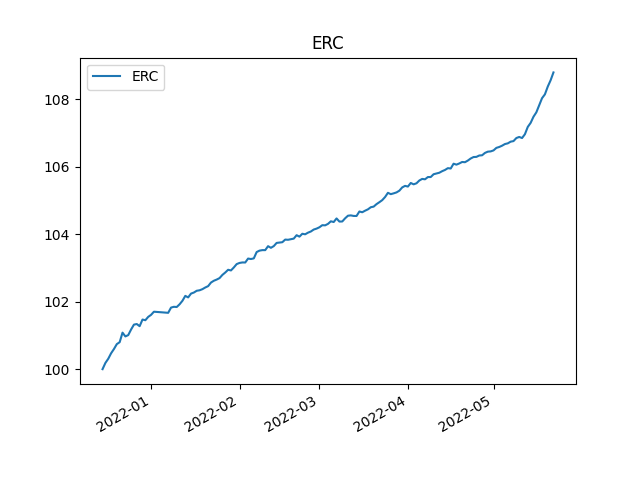}
    \caption{DeFi portfolio USD based}
    \label{fig:port_erc_usd}
\end{figure}

\begin{figure}[h!]
    \centering
    \includegraphics[width=1\columnwidth]{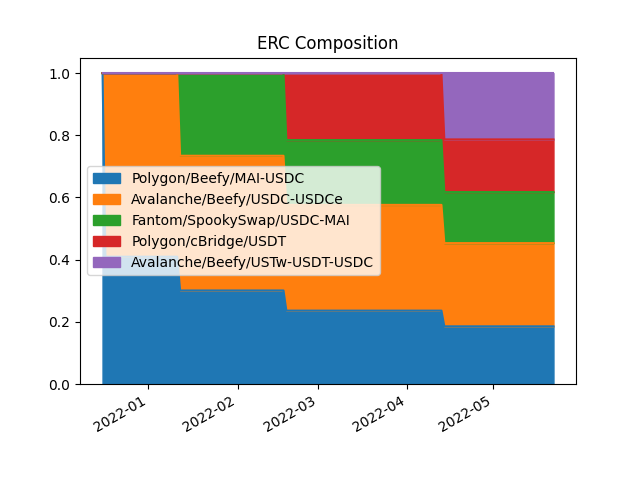}
    \caption{DeFi portfolio Composition}
    \label{fig:port_erc_compo}
\end{figure}

\subsection{The evolution of risk}

\begin{figure}[h!]
    \centering
    \includegraphics[width=1\columnwidth]{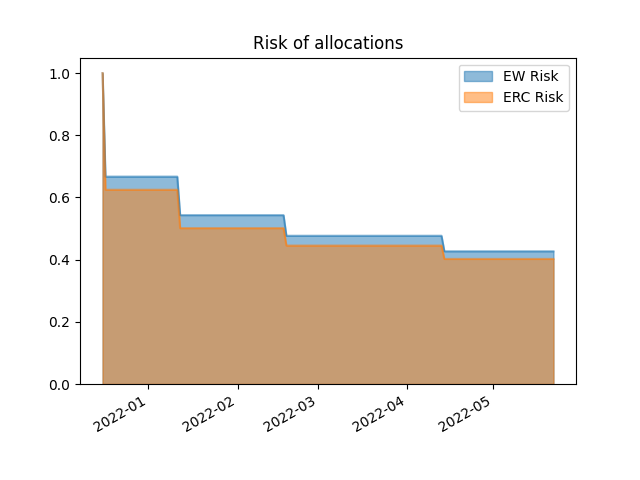}
    \caption{DeFi portfolio historical risk}
    \label{fig:port_risk_comp}
\end{figure}

As we rebalance the portfolio everyday, at each timestep the risk exposure is modified. For the simulation we started with two protocols, as the others were not deployed, and day after day some protocols are added to the allocation. Fig. \ref{fig:port_risk_comp} shows the evolution of risk following the addition of new protocols.

\subsection{Results}

We compared both DeFi portfolio allocations in order to assess the contribution of applying a scoring methodology on our portfolio. The comparison is based on a ratio performance over the risk of the allocation.  

\begin{figure}[h!]
    \centering
    \includegraphics[width=1\columnwidth]{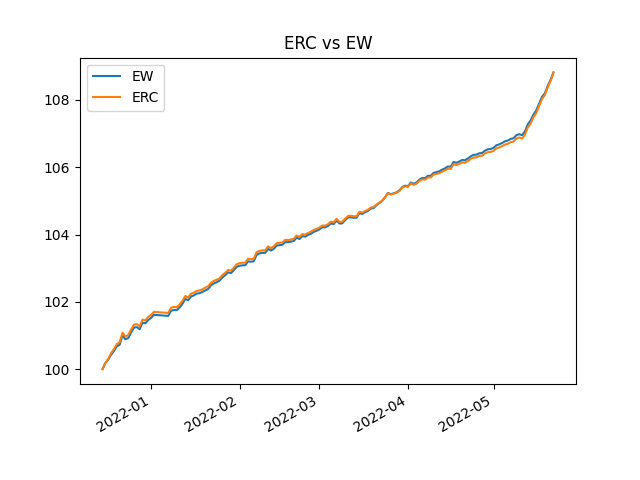}
    \caption{DeFi Portfolio ERC vs EW (USD based)}
    \label{fig:comp_backtest}
\end{figure}

\begin{table}[h!]
\centering
\begin{tabular}{lll}
           & EW       & ERC      \\\hline
2021-12-31 & 1.4400\% & 1.5290\% \\\hline
2022-01-31 & 1.5250\% & 1.5110\% \\\hline
2022-02-28 & 1.0720\% & 1.0520\% \\\hline
2022-03-31 & 1.2490\% & 1.1750\% \\\hline
2022-04-30 & 1.1180\% & 1.0550\% \\\hline
2022-05-22 & 2.1140\% & 2.1780\% \\\hline
\end{tabular}
\caption{Monthly performance of portfolios}
\label{table:monthly_perf}
\end{table}

\begin{figure}[h!]
    \centering
    \includegraphics[width=1\columnwidth]{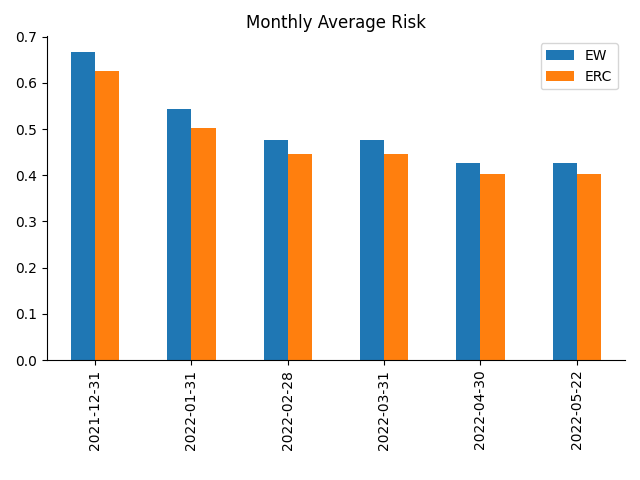}
    \caption{DeFi Portfolio ERC vs EW (USD based)}
    \label{fig:comp_backtest_risk}
\end{figure}

\begin{table}[h!]
\centering
\begin{tabular}{lll}
           & EW     & ERC    \\\hline
2021-12-31 & 0.6673 & 0.6249 \\\hline
2022-01-31 & 0.5433 & 0.5015 \\\hline
2022-02-28 & 0.4770 & 0.4455 \\\hline
2022-03-31 & 0.4770 & 0.4455 \\\hline
2022-04-30 & 0.4271 & 0.4024 \\\hline
2022-05-22 & 0.4271 & 0.4024\\\hline
\end{tabular}
\caption{Monthly Average Risk of portfolios}
\label{table:monthly_avg_risk}
\end{table}

\begin{table}[h!]
\centering
\begin{tabular}{lll}

           & EW     & ERC    \\\hline
2021-12-31 & 0,0216 & 0,0245 \\\hline
2022-01-31 & 0,0281 & 0,0301 \\\hline
2022-02-28 & 0,0225 & 0,0236 \\\hline
2022-03-31 & 0,0262 & 0,0264 \\\hline
2022-04-30 & 0,0262 & 0,0262 \\\hline
2022-05-22 & 0,0495 & 0,0541
\end{tabular}
\caption{Monthly Perf/Risk of portfolios}
\label{table:monthly_ratio}
\end{table}

When we look at Fig. \ref{fig:comp_backtest} we first notice that is difficult to make a difference between both portfolios. The statistics table \ref{table:monthly_perf} shows the performances are pretty similar, but the risk allocation is not the same for both porfolios observed in Fig. \ref{fig:port_risk_comp}.

\begin{figure}[h!]
    \centering
    \includegraphics[width=1\columnwidth]{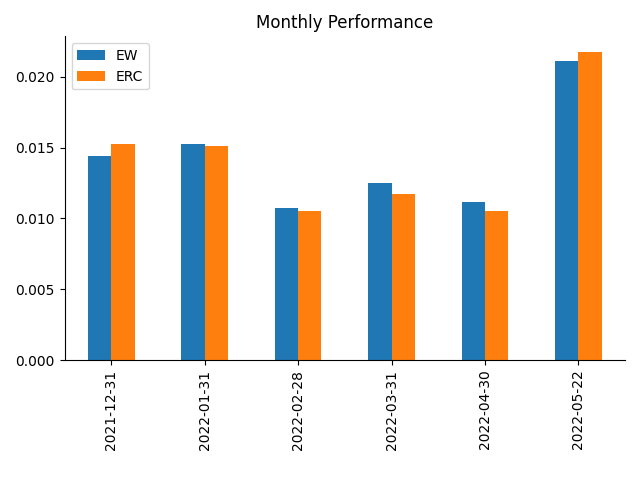}
    \caption{DeFi Portfolio ERC vs EW monthly performance}
    \label{fig:comp_monthly_perf}
\end{figure}

\section{Conclusion}
When we look at the portfolios that are equal-weighted and equal-risk-contribution-weighted, the yields obtained are pretty similar. However, Fig. \ref{fig:port_risk_comp} shows the difference in risk between the two portfolios and table \ref{table:monthly_ratio} shows a slight difference between both methodologies. The equal risk contribution allocation outperforms the equal weight methodology when we look at the ratio performance over risk.

The methodology we propose is similar to the approach of traditional asset managers. Nevertheless, using decentralized investment vehicles brings new structural risks to allocation. These risks are quantified by analytic providers, and in addition to this, the aggregated scores might fluctuate according to the views of the risk manager. We developed a scoring methodology to combat this. 

As a future work, it would be interesting to add views to the raw scores as well as introducing dependency between protocols established on the same blockchain. These views could not only be based on quantifiable criteria but also on a fundamental approach of the DeFi project. It would also be interesting to develop risk scores based on machine learning techniques, scoring methods, and decision trees. These methodologies could be a guideline for the  view matrix construction. It will also allow the measurement of the risk of each decentralized investment vehicle in real time and the possibility to build risk metrics specific to these new investment vehicles.

\bibliographystyle{IEEEtran}
\bibliography{bib}
\nocite{*}

\end{document}